# Can AI and humans genuinely communicate?

Constant Bonard

## Abstract

Can AI and humans genuinely communicate? In this article, after giving some background and motivating my proposal (§1–3), I explore a way to answer this question that I call the 'mental-behavioral methodology' (§4–5). This methodology follows the following three steps: First, spell out what mental capacities are sufficient for human communication (as opposed to communication more generally). Second, spell out the experimental paradigms required to test whether a behavior exhibits these capacities. Third, apply or adapt these paradigms to test whether an AI displays the relevant behaviors. If the first two steps are successfully completed, and if the AI passes the tests with human-like results, this constitutes evidence that this AI and humans can genuinely communicate. This mental-behavioral methodology has the advantage that we don't need to understand the workings of black-box algorithms, such as standard deep neural networks. This is comparable to the fact that we don't need to understand how human brains work to know that humans can genuinely communicate. This methodology also has its disadvantages and I will discuss some of them (§6).

**Constant Bonard**

# Can AI and humans genuinely communicate?

## 1. ELIZA

ELIZA was a computer program designed in the 1960s by Joseph Weizenbaum with the purpose of conversing with humans (Weizenbaum, 1966). It was one of the first programs that could take the Turing test.[1] In a version of the program, ELIZA takes the role of a psychiatrist. Here is a representative sample of a conversation that users had with it, where the capitalized utterances are ELIZA's:

> Men are all alike.
> IN WHAT WAY?
> They're always bugging us about something or other.
> CAN YOU THINK OF A SPECIFIC EXAMPLE?
> Well, my boyfriend made me come here.
> YOUR BOYFRIEND MADE YOU COME HERE
> He says I'm depressed much of the time.
> I AM SORRY TO HEAR YOU ARE DEPRESSED
> It's true. I am unhappy.
> DO YOU THINK COMING HERE WILL HELP YOU NOT TO BE UNHAPPY?
> I need some help, that much seems certain.
> WHAT WOULD IT MEAN IF YOU GOT SOME HELP?
> Perhaps I could learn to get along with my mother.
> TELL ME MORE ABOUT YOUR FAMILY
> My mother takes care of me.
> WHO ELSE IN YOUR FAMILY TAKES CARE OF YOU?
> My father.
> YOUR FATHER.
> You are like my father in some ways.
> WHAT RESEMBLANCE DO YOU SEE?
> (Weizenbaum, 1966, pp. 23–24)

ELIZA appeared to many of its users as being able to understand them, reflect upon the issues they raised, and give thoughtful answers – just like a competent psychiatrist would. But this was an illusion. ELIZA couldn't understand its users, nor could it genuinely communicate about the issues they raised. This can be shown by analyzing its code, as was pointed out by ELIZA's creator (Weizenbaum, 1966; 1976) and other scholars (Block, 1981; Shah et al., 2016). The simplicity of this program (200 lines in BASIC) makes it easy to show the tricks it uses to create the illusion of genuine communication. Let me describe some of them.

---

[1] A test where, roughly, a human participant converses with either another human or a computer and that is successfully passed by the computer if the participant cannot tell which one it is (Turing, 1950).

A first trick is to look for keywords in the users' prompts and to respond with a set of predetermined answers. When it spots the words 'mother,' it will give one of its mother-responses, such as 'Tell me more about your family'. If it spots the keywords 'depressed' or 'unhappy,' it will say, 'I am sorry to hear you are depressed/unhappy' or 'Do you think coming here will help you not to be depressed/unhappy'? A second trick is to repeat phrases of the user by preceding them with 'What makes you think that …' or 'Does it please you to believe …' while using simple grammatical transformations such as substituting 'I' with 'you' and 'my' with 'your.' A third trick is used when the program doesn't spot keywords in a sentence: It ignores the sentence and comes back to something the user said previously. A fourth (kind of 'meta') trick is the fact that ELIZA mimics the role of a Rogerian psychotherapist. As Weizenbaum explains, a psychiatrist appears as legitimate when asking questions that may, in another context, seem naïve or irrelevant. For instance, if a psychiatrist responds to a boat story with 'Tell me more about boats,' one wouldn't assume that they know nothing about boats (Weizenbaum, 1966, p. 26).

Weizenbaum has always been very explicit that ELIZA cannot genuinely converse with its users as a psychiatrist can. He was thus very surprised when he realized that many people who had the opportunity to know how the program works nevertheless attributed feelings or the capacity to understand them to ELIZA:

> I was startled to see how quickly and how very deeply people conversing with [ELIZA] became emotionally involved with the computer and how unequivocally they anthropomorphized it. Once my secretary, who had watched me work on the program for many months […] started conversing with it. After only a few interchanges with it, she asked me to leave the room. [This suggested that] extremely short exposures to a relatively simple computer program could induce powerful delusional thinking in quite normal people. […] This reaction to ELIZA showed me more vividly than anything I had seen hitherto the enormously exaggerated attributions an even well-educated audience is capable of making, even strives to make, to a technology it does not understand. (Weizenbaum, 1976, pp. 53–58)

Here is the moral I draw from ELIZA: People interacting with an AI may easily think that it can genuinely communicate with them – perhaps because we have a natural tendency to anthropomorphize certain

behaviors – but a deeper knowledge of how this AI works makes it clear that they were fooled (cf. the Clever Hans case, Samhita & Gross, 2013).

ELIZA's deceitful tricks are easy to point to. However, with more recent AIs – and typically with Large Language Models (LLMs) such as BERT, LaMDA, GPT, or Gemini – the matter is more complicated. And it will probably be even more so in the coming years. This is so for two reasons. First, the limitations of AI communicative abilities decrease rapidly over time. The launch of ChatGPT in 2022 made that strikingly clear to the public worldwide. Second, unlike ELIZA, most AIs today are based on artificial neural networks that are 'black boxes': The interactions between the many layers of artificial neurons that compose them are so complex that no one understands the dynamics of these networks (Beisbart & Räz, 2022).

Despite these two important differences, it is helpful to start with ELIZA's example to think about the following questions: What grounds the judgment that we can genuinely communicate with a psychiatrist but that ELIZA only mimics such an interaction? How to explain this when a psychiatrist and ELIZA may use the same words in response to our utterances? Why think that ELIZA tricks its users and that the impression that it understands them is an illusion?

There is certainly no universally accepted answer to these questions. They touch on deep issues concerning the nature of communication that are actively debated today. However, the philosopher of language Paul Grice and his heirs have proposed relevant hypotheses that are now widely accepted in various fields. In the following, I will discuss the possibility of human-AI communication based on these hypotheses.

Before, let me highlight that the issue of whether humans and AI can genuinely communicate is important for several reasons. From a theoretical perspective, investigating it should shed new light on theories of communication, meaning, and understanding – just like comparing human and nonhuman animal communication has (see, e.g., Moore, 2018; Tomasello, 2008). But this issue is also important from societal and ethical perspectives. AI will play more substantive and critical communicative roles in the future – as artificial colleagues, playmates, or, indeed, therapists. If we misplace an overly great trust in the communicative abilities of such AIs, this may lead to a range of harms, notably due to false, immoral, or irrelevant information (Bender et al., 2021; Henderson et al., 2018; Kasirzadeh & Gabriel, 2023; Weidinger et al., 2022). It is thus important to understand well the nature of these communicative interactions to avoid anthropomorphism and other forms of misinterpretation. Another reason concerns 'AI alignment': the problem of whether future, more independent AIs will possess goals that are in alignment with human values and norms (Bostrom & Yudkowsky, 2018; Liao, 2020). As Peter Railton suggested, successful human-AI communication may be a key contribution to managing this important issue because it may be an important stepping stone towards AI with a capacity for ethical learning (Railton, 2020).

## 2. What is communication?

It is trivial that AI and humans can communicate if 'communication' is understood in a broad sense, as is common in biology, to mean an exchange of information where both the sending and the receiving of that information were designed for that purpose (see, e.g., Hauser 1996, Scott-Phillips 2008). In this sense, I communicated with any chatbot when it correctly answered one of my questions.

But what linguists and philosophers usually mean by 'communication,' whether verbal or non-verbal, is more restrictive. They usually follow Grice (1989) in using this word to refer to an interaction requiring the *recognition of overtly displayed communicative intents.* These are cases where participants in the interaction know, and mutually know they know, that an attempt at communication has taken place. Grice and his heirs arrived at this circumscribed notion of communication notably by contrasting scenarios such as the following (Grice, 1957, pp. 381–382).[2]

- Scenario 1: Ludo leaves Merel's handkerchief near the scene of a murder to make the detective believe that Merel was the murderer. Ludo succeeds; his intention to create the relevant belief in the detective is fulfilled. But that doesn't count as a case of genuine communication. The reason is that Ludo *hides* his intention to produce a belief in the detective.
- Scenario 2: Ludo and the detective are on the crime scene, and the detective asks who could be the murderer. Instead of speaking, Ludo looks the detective in the eyes, takes out what they both know is Merel's handkerchief, and theatrically throws it on the crime scene. As long as he manages to make it mutually known that he intends to produce the belief 'Merel is the murderer,' Ludo would have successfully communicated this piece of information.

The relevant difference here is that, in the second scenario, Ludo not only had the intention to produce the belief that Merel is the murderer but also made sure that this intention was *wholly overt* so that the detective could recognize it. It is the recognition of overt intentions that makes an interaction count as genuine communication – it doesn't matter whether it is done through speaking, writing, drawing, gesturing, or throwing a handkerchief. Cases where overt intentions to communicate are recognized are often called 'ostensive communication.'

---

2   In this article, Grice is concerned with characterizing two sense of 'meaning', not with defining communication. However, he establishes a correspondence between 'non-natural meaning' and communication (Grice, 1957, p. 380; Grice, 1989, p. 291 or 367).

Now, many scholars have hypothesized that distinctively human communication is ostensive – whether it is verbal, gestural, pictorial, etc. (Bach & Harnish, 1979; Green, 2007; Grice, 1989; Heintz & Scott-Phillips, 2023; Recanati, 2019; Sperber & Wilson, 1986; Tomasello, 2008). Besides the comparison of contrasting cases such as scenarios 1 and 2, there is another, more fundamental, reason for this hypothesis: The ostensiveness of human communication seems to be required to explain its impressive expressive richness.[3] Let me briefly explain.

Famously, Grice (1975) proposed that once we recognize an intention to communicate, and in particular when we are engaged in a conversation, we expect people to respect pragmatic principles that he called 'conversational maxims': to be truthful, informative, relevant, and clear. For Grice, it is because we expect people to respect these principles that we can infer what people mean beyond what words literally mean. If, for instance, Adrien asks Pauline 'Did Joe bring some beer to the party?' and she answers 'He forgot his wallet,' Adrien will expect Pauline's answer to be relevant and informative and, based on these expectations, infer that Joe didn't bring beer. A knowledge of English – of what is *encoded* or *literally* expressed in this sentence – isn't sufficient to understand what Pauline means.

Grice's heirs have diverted from the original theory in various ways, notably by proposing their own pragmatic principles *(*Horn, 1984; Sperber & Wilson 1986; Levinson, 2000; Roberts, 2018). The general idea, however, has remained the same: The display and recognition of overt communicative intentions ground the communicators' mutual expectations that they behave by following certain principles such as Grice's maxims. These principles, in turn, explain humans' ability to mean more with words and other signs than what is encoded in these signs. Thus, although the encoded meaning of ostensive signs (e.g., the literal meaning of the words used by Pauline) is insufficient to account for what is communicated with these signs, the fact that they are ostensive, together with expectations that people usually respect the principles, can explain the distinctive expressive richness of human communication (for reviews see Korta & Perry, 2020; Horn & Ward, 2004; Schlenker, 2016; Bonard, 2021b, chap. 1).

---

3 These claims are not universally accepted (see, e.g., Millikan, 2004). Even within a Gricean perspective, there are cases of "subostensive" communication that seem to require uniquely human cognitive capacities (Bonard, 2022, 2023b). And nonhuman communication may well be more expressively rich than scholars have long thought (Bonard, 2023a; Schlenker et al., 2016).

Now, to communicate ostensively, philosophers and linguists have hypothesized that humans must possess various mental states and mechanisms. As we just saw, Grice (1989) proposed that the interlocutors must at least *recognize* each other's *communicative intentions* and have certain *expectations* concerning how people behave. Grice's heirs have refined such hypotheses, e.g., by requiring communicators to share a *common knowledge* (Lewis, 1969, 56), to make *presuppositions* (Stalnaker, 1978), or to detect what is *mutually manifest* (Sperber & Wilson, 1986, 42).

Following this line of thought, if we trust standard contemporary pragmatics, to know whether an AI can genuinely communicate with humans – in the substantive sense of ostensive communication – we need to know whether this AI possesses features sufficiently similar to the mental states and mechanisms targeted by notions like 'communicative intention,' 'presupposition,' or 'common knowledge.' I will call 'ostensive abilities' the set of mental abilities that are required for ostensive communication (according to a given theory), whether these abilities are instantiated by a brain or a computer. Now, how can we know whether an AI has these ostensive abilities?[4]

Clearly, some chatbots don't. ELIZA, for example, cannot recognize communicative intentions or expect people to behave by following pragmatic principles. However, the issue becomes less clear-cut with recent AIs, notably LLMs. To go back to a previous example, when I asked ChatGPT-3.5 what Pauline meant by answering 'He forgot his wallet' to Adrien, it correctly replied that Joe probably didn't bring beer to the party. Does that mean that it has what it takes to communicate ostensively? What is your intuition as a reader?

Some of you may be tempted to answer positively. However, let us not forget that we may be fooled by sophisticated tricks. As the creator of ELIZA warned us, we should beware of exaggerated attributions and anthropomorphism (Weizenbaum, 1976, p. 58).

Some of you, instead, may be tempted by a negative answer, perhaps because current LLMs sometimes display odd behaviors that reveal their incapacity to understand what we mean (ChatGPT may have gotten lucky with the beer example). However, let us remember that the progress made in the last couple of years is breathtaking. Surely, AI will continue to improve to a point where it will become impossible to distinguish their conversational skills from those of humans solely based on the subjective impressions given by their apparent conversational skills.[5] And, at this point, if the refusal to attribute ostensive abilities to AIs endures, this may be due to a chauvinistic attitude that draws an arbitrary barrier between humans and AIs.

---

4  This echoes classic questions in the philosophy of AI (see, e.g., Block, 1981; Dretske, 1985; Lucas, 1961; McCarthy & Hayes, 1981; Putnam, 1960; Searle, 1980). However, developments in AI and in pragmatics require new inquiries.

When AIs become fluent conversers, our intuitions are threatened by anthropomorphism and chauvinism. It is not reliable to just guess whether they have the required ostensive abilities based on everyday interactions with them. Crucially, as mentioned above, whether AIs can genuinely communicate is important, not just for theoretical reasons. It is also an important ethical and societal question. So, are there more reliable methodologies?

In section 4, I will propose what I call the 'mental-behavioral methodology.' But first, let me briefly discuss an alternative that I call the 'algorithm analysis methodology' and explain why I consider it to be insufficient for our purpose.

### 3. The algorithm analysis methodology and the black-box problem

The algorithm analysis methodology consists in analyzing the features of the algorithm of the candidate AI and assessing whether these features can correspond to those defining ostensive abilities. As I illustrated above, it is the methodology used by Weizenbaum (1966, 1976) for ELIZA: he analyzes the simple 200-line code he created to make it apparent that its tricks don't amount to having ostensive abilities. This example shows that the algorithm analysis methodology is very convincing when it is successful.

The problem is that the algorithms used by recent AI cannot be so easily analyzed. This is what I call the *black-box problem.* Contrary to traditional rule-based algorithms like ELIZA (or to successful 'explainable AI' (Adadi & Berrada, 2018)), many recent AIs are based on deep learning algorithms that are black boxes: the interactions between the many layers of artificial neurons that compose them are so complex that, even when computer scientists know their parameters and connections, they still can't understand the dynamics of the network (Beisbart & Räz, 2022). This is both because the computation paths from inputs to outputs have many steps – they have numerous layers of artificial neurons – and because their designers don't specify their computational steps but rather how the AI 'learns' to make predictions based on training datasets (Russell & Norvig, 2021, Chapter 22).

---

5  Several studies already show that humans are not able to distinguish with certainty between human-created and machine-generated text (e.g. Clark et al. 2021, Schwitzgebel et al. 2023).

Let me briefly illustrate the limitations of the algorithm analysis methodology with two examples. Landgrebe and Smith (2021) use this methodology to argue that *transformers* – the type of deep learning architecture used by, e.g., ChatGPT – lack the semantic abilities required for genuine communication. They start from the premise that humans can learn languages only because they possess a large body of knowledge before starting to acquire the grammar of their mother tongue. They claim, for instance, that language acquisition presupposes knowing the distinctions between objects and processes, individuals and categories, or natural and accidental properties. Now, if we analyze the algorithm used in transformers, we realize that they lack any such knowledge before their training phase. Transformers are *agnostic algorithms* insofar as they do not rely on any prior knowledge about the task or about the types of situations in which the task is performed before they are trained on datasets. For this reason (among others), Landgrebe and Smith assert that the vector space that transformers use is merely morpho-syntactical and lacks the necessary semantic dimensions for genuine communication.

Here is my worry about this reasoning. Let us assume, for the sake of the argument, that Landgrebe and Smith are correct about the following claims: (a) humans possess a large body of knowledge (about objects, processes, and so on) before acquiring language, (b) humans require this knowledge to acquire language, (c) this knowledge is not encoded in agnostic algorithms like transformers, (d) this knowledge is required by any agents to correctly interpret human language and so to genuinely communicate with them. It doesn't follow from (a)–(d) that transformers cannot genuinely communicate with humans because it doesn't follow that the relevant body of knowledge must be possessed by all language learners *before* acquiring language – (b) only states that *humans* must. It may be the case that the relevant knowledge is required (as per (d)) but that transformers acquire it when they are trained with an adequate dataset. In other words, although transformers learn languages in ways that are very different from how humans learn them, transformers may nevertheless end up with the adequate knowledge to correctly interpret human language. Thus, the argumentation from the last paragraph doesn't show that transformers cannot acquire the semantic abilities required for genuine communication. The question then is: how would we know whether they can? The problem with the algorithm analysis methodology is that it seems that this cannot be known by analyzing transformers' algorithms since they are black boxes.

Here is another short example of the algorithm analysis methodology: Chalmers (2020) remarked that GPT-3 is 'not much of an agent' because 'it does not seem to have goals or preferences beyond completing text.' This may suggest that GPT-3 can't communicate ostensively since ostensive communication requires other goals, in particular, the goal of inferring communicative intentions. Similarly, Bultin and Viebahn (2023) argue that PaLM can't make genuine assertions because (roughly) it doesn't have the goal of conveying information but merely to select for outputs that contain likely words.

However, even though GPT-3's and PaLM's algorithms initially only have the goal of completing text or selecting for outputs that contain likely words, this doesn't exclude that, during their training phase, these models develop instrumental intermediary goals to achieve the initial one. It is at least possible that the 175 billion parameters of GPT-3, after being trained on 45TB of text data, form a network that, functionally speaking, includes the intermediary goal of inferring the communicative intention of the text producers, perhaps because it is the easier route for it to achieve its initial goal of completing their text. I am not claiming that this is what happens, but how could we rule out this possibility by analyzing the algorithm? I don't think we can, and thus, it is at least in principle possible that GPT-3 (or other such models) develops the goals necessary for ostensive communication during its training phase, even though these goals weren't coded in its initial algorithm. After all, the neuronal networks of many animal species – their brains – plausibly create rather sophisticated intermediary goals during their 'learning phases' as they discover their environment in order to achieve initial goals that may be quite basic (e.g., maximize the expected cumulative reward, as some reinforcement learning theories predict, see Sutton & Barto, 2018).

In the two examples I have given, the black box problem means that the algorithm analysis methodology is insufficient to exclude the possibility that the AIs in question have acquired the relevant ostensive abilities during their training phase. Perhaps one day, computer scientists will be able to successfully analyze these black boxes and tell us whether or not these AIs learn the relevant knowledge and goals. But until then, it appears that the algorithm analysis methodology is insufficient to answer the question we started with for all the currently best models.

Let me now turn to my positive proposal.

## 4. The mental-behavioral methodology

In this section, I will present the mental-behavioral methodology – MBM for short. The MBM has a major advantage over the algorithm analysis methodology: it doesn't require that we understand how black-box algorithms work. This is comparable to the fact that we don't need to understand how human brains work to know that humans can genuinely communicate. The MBM is not new: as we will see in the next section, some researchers who study human-AI communication can be said to already engage in it. However, the MBM hasn't, as far as I know, been formulated or discussed as such.

At a general level, the MBM consists of three steps and an evaluation rule:

- First step: Spell out what mental capacities are required for human communication (as opposed to communication more generally) given the best theory.
- Second step: Spell out or develop the experimental paradigms required to test whether a behavior exhibits these capacities.
- Third step: Apply or adapt these paradigms to test whether an AI displays the relevant behaviors.
- Evaluation rule: If the first two steps are successfully completed, and if the AI passes the tests with human-equivalent results, this constitutes evidence that this AI and humans can genuinely communicate.

Here are two observations about the practicability of the MBM.

First, the MBM can either go it alone or work in combination with other methodologies. In the first alternative, the evaluation rule is the sole source of evidence. In the second alternative, which is certainly wiser, the evidence from the evaluation rule should be weighted with evidence from other sources to evaluate the possibility of genuine human-AI communication. The MBM, however, is silent about what other sources of evidence there are and their comparative weight. In the following, I will only consider the "go it alone" alternative.

Second, the MBM cannot yet be fully completed because the science required is itself not complete: there are important disagreements among theories of human communication and the mental capacities postulated by these theories cannot all be measured in behavioral tests. Nevertheless, the theories and experimental paradigms that exist can already take the MBM pretty far – even if it can never be fully completed. They permit an accumulation of evidence approach whereby the ability of an AI to pass a variety of tests adds weight to the claim that this AI is capable of genuine communication (such an approach has been proposed, for instance, in research on nonhuman animal consciousness).[6]

To evaluate the plausibility of the MBM, let us consider a strong version that puts aside these two practical matters – i.e., that evidence can come from other methodologies and that it can accumulate in a gradual manner as the science progresses. The strong version of the MBM can be formulated as such:

*strong MBM hypothesis*
- If the $AI_X$ successfully passes the behavioral tests $B_Y$ for the set of cognitive capacities $C_Y$, then $AI_X$ has the ability to genuinely communicate with humans, according to the theory of communication $T_Y$ and given the communication fragment $F_Y$.

Where:
- $AI_X$ is the candidate AI for human-AI communication. Examples are ELIZA, Gemini, GPT-3, etc.
- $C_Y$ is the set of cognitive capacities that are postulated as necessary and sufficient by the theory $T_Y$ to genuinely communicate. A non-exhaustive list of candidate elements includes communicative intentions, beliefs, assumptions, expectations, and a Theory of Mind (i.e., a capacity to ascribe the appropriate mental states to others).
- $B_Y$ is a set of behavioral tests such that, when they are successfully passed, show that the human or the AI that took them possesses the set of cognitive capacities $C_Y$ (to a certain degree). Examples include the behavioral tests designed to assess an agent's Theory of Mind that can be taken by AIs (I will discuss this in section 5 below).
- $T_Y$ is a theory of communication from cognitive science that makes a hypothesis of the following form: 'To be able to genuinely communicate (within the communicative fragment $T_Y$), it is necessary and sufficient to have the set of cognitive capacities $C_Y$.' Potential examples include relevance theory and cog-sci versions of various pragmatic theories (e.g., Grice's, Stalnaker's, or other speech act theories, see Fogal et al. (2018)).
- $F_Y$ is a fragment of communication that corresponds to theory $T_Y$'s scope, i.e., over which $T_Y$ theorizes. Possible examples include a fragment of written English, oral Franco-Provençal dialects, road signs, or facial expressions.

---

6   Thanks to Paula Droege for pointing this out to me.

If the strong MBM hypothesis is true, the MBM should allow answering the question of this paper – whether humans and AI can genuinely communicate – with a degree of certainty proportional to the degree of certainty that a theory $T_Y$ is correct and that the tests $B_Y$ are appropriate to $T_Y$. In other words, the more confident we are about $T_Y$ and $B_Y$, the more confident we should be about the answer to our question if the strong MBM hypothesis is true.

Putting aside for the moment whether we can arrive at a correct $T_Y$ and appropriate $B_Y$, is the strong MBM hypothesis plausible at all? In other words, roughly, can we really know whether two agents are able to genuinely communicate by behavioral tests of their cognitive capacities?

To answer this question, I need to do a digression concerning fundamental theories in the philosophy of mind because I believe that the MBM presupposes a form of functionalism. Functionalism is the theory that a type of mental state can be identified by its function to cause and to be caused by certain changes in other states of the subject, e.g., something is fear if it has the function to cause certain states (e.g., avoidance) and be caused by certain states (e.g., appraisals of imminent danger). The dependence of the MBM on functionalism may be made clearer by contrasting the MBM with other methodologies stemming from rivals to functionalism: from forms of (a) logical behaviorism, (b) type physicalism, and (c) psychological phenomenalism.

(a) Logical behaviorism says that a mental state should be defined solely in terms of behavioral dispositions. To assess whether humans and AI can genuinely communicate, a behaviorist methodology may use a Turing test: If human evaluators cannot reliably tell whether they are conversing with an AI or with another human (given a certain context, amount of time, etc.), then the AI would be able to genuinely communicate. This purely behaviorist methodology does not factor in the mental capacities grounding communication; it focuses solely on the observable behavior. By contrast, the MBM is not behaviorist because, to borrow an expression from Block (1981, p. 36), it indulges in 'a kind of theorizing about cognitive mechanisms that would be unacceptable to a behaviorist.' MBM is functionalist insofar as it purports to identify mental capacities – typically, ostensive abilities – based on their functions to cause and be caused by other states and where these functions are revealed through behavioral tests, e.g., tests that are supposed to measure whether a subject's Theory of Mind functions properly by attributing the appropriate mental states in others (I will discuss such tests in section 5).

(b) Type physicalism says that a type of mental state should be identified with the type of physical state that correlates with it. According to it, the mental states making up genuine communication are identified with brain states. To assess whether AI-human communication can be real, a type-physicalist methodology would assess whether the hardware making up an AI is sufficiently similar to human brains. By contrast, the MBM is open as to whether the physical realization of an AI must resemble human brains. It is not completely indifferent to this issue insofar as showing that an artificial neural network behaves similarly to human brains may be evidence that the AI and humans treat stimuli by using the same functions (more on this below). This example illustrates that 'behavior' in the MBM should be understood broadly to include the behavior of brains and artificial neural networks. This provides a connection between type-physicalism and the MBM. Note, however, that a given behavior of artificial neural networks can typically be instantiated by different types of hardware and so can have multiple physical realizations. Like functionalism and unlike type physicalism, the MBM embraces this possibility.

(c) Psychological phenomenalism says that a type of mental state should be identified by the type of subjective feeling that is associated with it, i.e., by its phenomenal character. Thus, according to it, the states making up genuine communication may be identified by the way they feel, e.g., by what it is like to grasp that someone intends to communicate. Until we have some evidence about AI feelings, I don't see how this theory could lead to a methodology that would help us answer our question. By contrast, the MBM would allow for the possibility that an AI can genuinely communicate with humans even if we don't ascribe feelings to it. However, it is also possible that phenomenal consciousness plays an essential *functional role* in human communication, e.g., in understanding. In that case, the MBM may make predictions in line with psychological phenomenalism, at least concerning certain mental states required by communication. I will get back to this in section 6.

Fortunately for the MBM, although there are issues with functionalism, its advantages continue to attract philosophers. It appears to remain the most commonly accepted theory and, in particular, among philosophers of cognitive science (for consciousness, see Bourget et al., 2023; for belief, see Schwitzgebel, 2021). However, even if one accepts functionalism, there are many potential issues with the MBM. Before we turn to these and in order to have a better idea of what the MBM may amount to, let us put more flesh on the bones presented in this section.

## 5. Some tentative first steps for the mental-behavioral methodology

In this section, I will briefly sketch possible ways to make progress on the first two steps of the MBM. My goal is not to establish definite solutions but to give some indication that the MBM appears to be a feasible way forward and potentially a promising one for future research on human-AI communication.

### *5.1 First step*

As a reminder, the first step of the MBM is the following: Spell out what mental capacities are required for human communication. Following the broadly Gricean framework of this paper, I see six main questions that need to be answered for this step. They are more or less interconnected in the sense that responding to one requires at least partial responses to other questions, but this may vary from theory to theory. I also highlight some important subcategories for these questions. This list is certainly not exhaustive, but it should help to illustrate some of the challenges faced by the MBM.[7]

(1) What does it take for someone to mean something with a sign? Or, as this is often formulated, what does it take to *speaker-mean*?
  – What does it take to speaker-mean in the most minimal, less demanding cases? Can babies or nonhuman animals speaker-mean (see Moore, 2018)? Does an agent with the capacities for minimal speaker-meaning possess the ostensive abilities sufficient to master all kinds of speaker-meaning (see, e.g., the distinctions between illocutionary vs. factual vs. objectual speaker-meaning in Green, 2007)? If not, what further capacities must the agent possess?
(2) What does it take to master the pragmatic principles humans use in communication?
  – Which of Grice's maxims are essential for human communication? Do we need several principles (as suggested by, e.g., L. Horn, 1984; Levinson, 2000; Schlenker, 2016)? Alternatively, can we derive all pragmatic principles from a single, more fundamental

---

7  Of course, there already is much research on these questions (for a sample far from being exhaustive, see, e.g., Bender & Koller, 2020; Butlin, 2023; Butlin & Viebahn, 2023; Cappelen & Dever, 2021; DeVault et al., 2006; Freiman & Miller, 2020; Green & Michel, 2022; Kasirzadeh & Gabriel, 2023; Lake & Murphy, 2023; Piantadosi & Hill, 2022; Stone, 2005).

one (as suggested by, e.g., Kasher, 1982; Roberts, 2018; Sperber & Wilson, 1986)?

(3) What does it take to share a common ground in the broad sense of the expression (which includes what Stalnaker (1978, 2014) refers to, but also 'common knowledge/belief' (Lewis, 1969; Schiffer, 1972), or 'mutual cognitive environment' (Sperber & Wilson, 1986))?

(4) Given a communicative fragment (e.g., written English), what does it take to master its encoded meaning (also called 'literal' or 'semantic' meaning, see, e.g., Schlenker, 2016)?

  – In particular, what does it take to master the semantics and the syntax of a language? Do we use these capacities to disambiguate indexicals, anaphora, and demonstratives, or do these cases belong to the next question? Does truth-conditional semantics coupled with syntactic rules exhaust encoded meaning (for a negative answer, see, e.g., the use-conditional meaning framework Gutzmann, 2015)?

(5) What does it take to infer nonencoded meaning (also called 'pragmatic' or 'strengthen' meaning, see, e.g., Schlenker, 2016)?

  – Do we require the same capacities to infer pragmatic presuppositions (Stalnaker, 1974), conventional, conversational, or nonconversational implicatures (Grice, 1989; Potts, 2005)? Is this question redundant with questions 1–4 or are there mental capacities that they don't point to?

(6) Finally, a question that goes beyond the scope of theories of communication but that, I think, must be answered in response to most, if not all, questions 1–5: What does it take to master a concept (for a detailed investigation of concepts in minds and LLMs, see Lake & Murphy, 2023)?

  – Are there concepts that are required by any form of ostensive communication (e.g., the concept of intention, of belief, logical concepts)? Are there categories of concepts that require non-conceptual capacities to be learned, and if so, what are these capacities? Relevant candidates are perceptual concepts (red, high-pitched, bitter), feeling concepts (feeling overwhelmed, aroused, tired, hungry, sad), or embodied concepts (sick, athletic, agile, cramped). These concepts may be especially difficult for AI to master, but they may include most or even all concepts according to some (radical) views (see, e.g., Barsalou, 2008).

Each of these six questions is a rabbit hole: Their answers open up new, difficult, interesting questions. Take question (3): Most theories define 'common ground' through mental concepts such as beliefs, assumptions, or what is cognitively manifest. The obvious question then is: How do you define these mental concepts? The notion of belief, typically, is itself defined in multiple, rather different ways depending on who you ask, including within the functionalist tradition in which the MBM is anchored (see, e.g., the interpretationalism of Dennett, 1987; the representationalism of Mandelbaum, 2014; the dispositionalism of Schwitzgebel, 2002). And the definition of the relevant mental states has to be given by bearing in mind that the MBM aims to measure their presence through behavioral tests in both humans and AIs.

However, this rabbit-hole situation is not hopeless. Even if there is no universal agreement on how to define a communicative ability or how to measure what defines it, we may agree that certain behavioral tests bring strong evidence for the possession of this ability.[8] Concerning beliefs, for instance, although there is no consensual definition, there is a rather wide agreement concerning functional characteristics of beliefs, i.e., the causal relationships that a belief typically has with other mental and behavioral states such as with other beliefs, perception, desire, and action (see Schwitzgebel, 2021, sec. 1.4). These characteristics may serve as inspirations for behavioral tests. Thus, although questions (1)–(6) are difficult, I remain hopeful that we can design tests allowing us to accumulate evidence for certain answers and adding weight to their plausibility.

### *5.2 Second step*

As a reminder, the second step of the MBM is the following: Spell out or develop the experimental paradigms required to test whether a behavior exhibits the communicative capacities identified in the first step. Assuming that the six questions presented in the last section are representative of the required communicative abilities, then this step is supposed to gather empirical evidence concerning an AI capacity to (1) speaker-mean, (2) master the pragmatic principles, (3) share a common ground, (4) master the encoded meaning of a communication fragment, (5) infer nonencoded meaning, and (6) master the relevant concepts.

My goal here is not to present tests that answer these questions – I don't think there are any as of yet – but rather to briefly discuss some examples that have already been administered to AIs. This should help to flesh out what this second step could look like.

---

8  Compare: If someone has the highest score on TOEIC, IELTS, and TOEFL, this seems to be enough evidence that they are proficient in English, even in the absence of universally accepted definitions of English proficiency or of its various components, such as the mastery of conjugation, lexicon, reading, pronunciation, and so on.

First, it should be noted that the number and diversity of tests that are administrated to AIs are already enormous and are rapidly growing. It is common practice for AI researchers to study how AIs function by measuring their performance on benchmark data and protocols. There already exist hundreds of such tests (see, e.g., Gemini Team, Google, 2023; Hendrycks et al., 2021; Mitchell & Krakauer, 2023; Srivastava et al., 2023). Furthermore, because recent AIs can use multimodal inputs – Gemini can use a mixture of texts, pictures, videos, and audios – most tests that cognitive scientists designed for humans can now be taken by AI. For these reasons, my short review will by no means be representative of the state-of-the-art results and miles away from what can be done in the (immediate) future.

### 5.2.1 Theory of Mind

According to most theories, it is essential for communication to possess the capacity to correctly attribute mental states to others, typically beliefs and intentions – a capacity often called *Theory of Mind* or *ToM* for short. For Griceans, ToM is required to respond to questions (1), (3), and (5), at least. The False Belief Task is an experimental paradigm that is widely used to test ToM. Here is a version of it. Subjects are told a story about Sally, who puts her candy in a box and leaves the room. In her absence, another character moves the candy to a basket. When Sally returns, subjects are asked where Sally will look for her candy. Young children have difficulties with this question: they typically respond that Sally will look in the box. They fail to attribute to Sally a belief that they know to be false. Being able to pass this test is often interpreted as evidence that one has a developed ToM.

Do AIs have a human-like ToM? As of December 2023, it seems that a majority of researchers give a negative answer, typically based on poor results that AIs have at the False Belief Task (Aru et al., 2023; Ma et al., 2023; Shapira et al., 2023; Stojnić et al., 2023; Ullman, 2023). Some are slightly more optimistic but remain rather agnostic (Holterman & van Deemter, 2023; Mitchell & Krakauer, 2023; Trott et al., 2022), while a few claim that, at least for GPT-4, we find human-like results for the False Belief Task (Gandhi et al., 2023; Kosinski, 2023).

Now, it should be noted that ToM is not a monolithic phenomenon (see also the contribution 'What mindreading reveals about the mental lives of machines' by Stephen Butterfill in this volume). It encompasses the abilities to attribute beliefs, emotions, reasoning, intentions, and other mental states, and these abilities may come apart (Apperly & Butterfill, 2009; Langley et al., 2022). Accordingly, different tests beyond the False Belief Task should be used by the MBM. For a position paper that aims to take this into account for research on LLMs' ToM, see Ma et al. (2023).

### 5.2.2 Infer nonencoded meaning

As highlighted in section 2, part of what makes human communication particularly rich is our ability to infer meaning beyond what signs encode (see also question (5)). Typically, we effortlessly resolve implicatures, cases where one means something by saying something else. Above, I gave this example: Adrien asks Pauline 'Did Joe bring some beer to the party?' and she answers 'He forgot his wallet'. Can AIs infer, as we do, that Joe didn't bring beer?[9]

Ruis et al. (2023, p. 2) investigated this question in detail and found that, currently, implicature resolution is a challenging task for LLMs (including for GPT-4, though it performs much better than its rivals). They conclude that 'pragmatic understanding has not yet arisen from large-scale pre-training on its own.'

### 5.2.3 Concepts, world knowledge, and general cognitive abilities

BIG-bench (Srivastava et al., 2023) is a battery of tests consisting of 204 tasks, contributed by 450 authors, and including problems from linguistics, childhood development, math, common-sense reasoning, biology, physics, and beyond. Having tested various LLMs, including GPT-4, Srivastava et al. find that they 'perform poorly on BIG-bench relative to expert humans' (2023, p. 25). However, performance improves with model size (effective parameter count), so that we have reasons to be optimistic about future models. Relatedly, Gemini is allegedly the first model to outperform human experts on MMLU (Massive Multitask Language Understanding), a large set of tests on knowledge across various domains (law, biology, history, etc.), as well as on reasoning and reading comprehension (Gemini Team, Google, 2023).

But do these tests tell us something about the cognitive proneness of the AIs tested and help answer questions (4) and (6) above? Or do they only show an unintelligent ability to repeat the knowledge LLMs have been trained with? Closer to the spirit of the MBM, Binz and Schulz (2023) used a battery of canonical experiments from cognitive psychology to assess GPT-3's decision-making, information search, deliberation, and causal reasoning abilities. Using these established experiments has important advantages over the benchmarks and protocols typically used by AI researchers – such as the BIG-Bench or the MMLU – since cognitive psychologists have carefully designed and refined these experiments to detect various cognitive biases and to disentangle different ways of how a task can be solved.

---

9  When I tested ChatGPT-3.5, it did.

Binz and Schulz's overall assessment is lukewarm. On some tasks, GPT-3 was better than humans, but they also find that 'small perturbations to vignette-based tasks can lead GPT-3 vastly astray, that it shows no signatures of directed exploration, and that it fails miserably in a causal reasoning task.' (Binz & Schulz, 2023, p. 1). Beyond these results, this study shows how an AI can be treated like a regular participant in canonical psychological experiments. This is what is more important for the MBM. See also Dasgupta et al. (2023) for an application to LLMs of canonical reasoning tasks from psychology and Lake and Murphy (2023) for a detailed analysis, based on psychological theories, of whether LLMs can master concepts.

### 5.2.4 Similarities between LLM embeddings and brain behaviors

I want to mention a last study because of its interesting methodology. Instead of studying AIs based on their performances at behavioral tests, as in the preceding examples, Li et al. (2023) compared how LLMs and human brains represent the same words, which is relevant to questions (4) and (6). To obtain the brain representation for a specific word, they used fMRIs of people reading texts, creating vectors for a timeframe before and after the person in the scan read the word. For the LLMs representations, they used their word embeddings, i.e., vectors that encode the meaning of words in such a way that the words that are closer in the vector space are expected to be similar in meaning. They then compared the fMRI vectors and the word embeddings using various statistical methods and found a 'remarkable structural similarity' (Li et al., 2023, p. 2).

As far as the MBM is concerned, what I find interesting about this study is the possibility of extending the comparison of AI and human behaviors beyond verbal reports, which all previous examples used, to brain behaviors. Indeed, nothing obliges the MBM to be restricted to verbal reports or overt behaviors. As long as there is some common ground where it makes sense to compare AI and humans, the MBM should welcome comparisons of more fine-grained behavioral responses. Besides brain behaviors, we can also think of brain lesion methods, response times, psychophysiological monitoring (e.g., for energy consumption), or even eye-tracking. All these types of experimental paradigms can, in principle, be used to test whether a (physiological or overt) behavior exhibits the mental capacities required for human communication. As the study by Li et al. (2023) shows, with a little ingenuity, they may be adapted to also test AI behaviors.

## 6. Difficulties for the MBM

The short review of the literature in the preceding section illustrates the wide array of tests that are already used to test communicative abilities in AI. This is promising for the MBM. However, using such tests raises several difficulties, three of which I will now discuss.

### *6.1 Difficulty 1: Shortcut learning*

An AI sometimes seems to master a capacity – e.g., detecting cows in photos – but is then shown to fail miserably when tested on a slightly different dataset – e.g., with photos of cows whose background isn't grass. This phenomenon is called *shortcut learning*. Shortcuts are decision rules that perform well on standard benchmarks but fail to transfer to more challenging testing conditions (Geirhos et al., 2020). Shortcut learning is frequent in current AI and is sometimes difficult to detect. Here is an example directly relevant to our topic: BERT was optimistically hailed as having human-like results on SuperGLUE, a standard benchmark for language understanding, but was then shown to perform very poorly on a version of the test that used slightly different wordings. The problem was that BERT answered these tests by detecting spuriously correlated keywords rather than with humanlike abilities (Mitchell & Krakauer, 2023).

The problem for the MBM is that an AI can successfully pass the battery of tests B that is supposed to measure the cognitive capacities C while, in fact, using shortcuts and not having C. How can the MBM make sure that its tests elude this trap?

Various solutions have been proposed to avoid shortcuts (Geirhos et al., 2020):

– First, tasks should be carefully selected. For instance, tests should use datasets that are systematically different from the training datasets. One can also use datasets for tests where known shortcuts are removed.
– Second, results should be interpreted carefully. For instance, researchers should always beware of anthropocentrism and never attribute high-level abilities that can be adequately explained by shortcut learning (a version of Morgan's Canon).
– Third, the mechanisms tested should be studied in as much detail as possible. AIs use shortcuts because the latter constitutes the easiest solution to a learning task (as animals and humans do, see Geirhos et al., 2020). Shortcuts can be avoided by understanding what is easy to learn. But doing so requires a detailed knowledge of the mechanisms used to learn.

These three solutions are helpful heuristics to diminish shortcuts, but they don't address a deeper problem that shortcuts point to: the problem of inferring (mental) capacities from observing behaviors.

### 6.2 Difficulty 2: Behavioral tests don't reveal mental capacities

An AI may display behavior typical of intelligent beings while there is a difference in their information processing such that the AI doesn't possess the cognitive capacities of the intelligent being. ELIZA fooling its users or hard-to-detect shortcuts point to this possibility. A thought experiment adapted from Block (1981) illustrates it: Imagine a robot that is designed to act like someone called Val based on sophisticated psychological theories from the future. These theories have produced an enormous lookup table predicting how Val would act (outputs of the lookup table) based on what sounds Val hears (inputs of the lookup table). When the robot detects soundwaves corresponding to a sentence in English, it responds as Val would have based on the lookup table. Although its ability to converse would be undistinguishable from Val's, this robot shouldn't be qualified as intelligent or as having genuine communicative abilities.[10] It merely displays the intelligence of Val and of the scientists who built it. A main conclusion that Block draws from this and other similar examples is that behaviorism about intelligence is false: Whether a behavior is intelligent depends on the character of the *internal* information process that produces it, not on its behavioral dispositions.

Since the MBM uses behavioral tests, does Block's conclusion mean that the MBM is doomed? No. Because the MBM isn't behavioristic: Its behavioral tests are meant to reveal internal information processes. As such, and as emphasized above, the MBM stands in stark contrast with the Turing test since the latter is agnostic about internal information processes. Now, it is true that some behavioral tests that are supposed to reveal internal information processes nevertheless would not be able to tell apart Block's robot from an agent with genuine communicative abilities. For instance, the robot's lookup table may yield the appropriate responses to a False Belief Task.

What this shows, I think, is that the MBM's tests must target cognitive mechanisms described at a fine-grained level (while still being coarse-grained enough to be implementable by either humans or computers). A test targeting the mechanism that takes sounds as inputs and yields verbal behaviors as outputs is very coarse-grained. Such a test wouldn't be able to detect the difference between a normal brain mechanism and Block's robot. However, finer-grained tests targeting the cognitive mechanisms at work in between the sound inputs and the verbal outputs could detect the difference. And let us not forget that the behavioral tests at the disposition of the MBM can target very specific behaviors and, in fact, can even target various brain behaviors, as we saw above when I presented a study on the structural similarities between LLM and brain representations of words (Li et al., 2023).

---

10 Block highlights the resemblance of his thought experiment to Searle's Chinese Room, but he disagrees with Searle's conclusions (Block, 1981, n. 30).

### 6.3 Difficulty 3: What if communication requires consciousness?

The MBM, even if it uses extremely fine-grained tests, may not be able to discriminate whether an agent is conscious. Indeed, it is at least conceivable that there could be AI completely lacking consciousness but that behaves exactly as we do and whose cognitive mechanisms are indiscernible from ours, even when scrutinized by fine-grained tests (cf. Block's China brain thought experiment (1978) or philosophical zombies (Chalmers, 1996)). If so, a third difficulty for the MBM is the possibility that our best communication theories require conscious cognitive processes. For instance, Recanati's Availability principle requires that 'what is said must be consciously available to the interpreter' (2004, p. 17). Understanding what is said by someone, for Recanati, requires a conscious experience of what is said, an experience that he compares to conscious perception (idem). Besides Recanati's, there are other theories that require understanding to be conscious (e.g., Bourget, 2017; Pepperell, 2022; Searle, 1980).

At this point, we should distinguish access consciousness from phenomenal consciousness. A representation is access-conscious when 'it is broadcast for free use in reasoning and for direct 'rational' control of action (including reporting)' (Block, 2002, p. 208). If something is available to you in your reasoning, action, and verbal report, then (roughly) it is access-conscious. Access consciousness, thus, is defined in functional terms, through the thoughts and actions it allows. By contrast, phenomenal consciousness is defined by what it is like to have a conscious experience, by the way it feels to be in that state, how it appears to one from the inside.

Now, if our best theories of communication require a capacity for access consciousness, then this is not a problem for the MBM because access consciousness can, at least in principle, be empirically tested since it is defined functionally. It would be problematic for the MBM if our best theories of communication required *phenomenal* consciousness. Indeed, it is at least conceivable that we'd have no way of knowing whether an AI that passes all tests with human-like results would or wouldn't be phenomenally conscious.

But is it the case that our best theories of communication require phenomenal consciousness? This is not obvious. For instance, we may well formulate Recanati's Availability principle using only access consciousness without, it seems, distorting the thesis. Similarly, mutual manifestness (Sperber & Wilson, 1986), common ground (Grice, 1989; Stalnaker, 2014), or common knowledge (Lewis, 1969; Schiffer, 1972), though they arguably require consciousness, are all defined in functional terms.

Nevertheless, there are some intuitive reasons to maintain that phenomenal consciousness is required, at least in certain domains of communication. I'm thinking in particular about emotionally loaded exchanges. Take a person who is looking for some empathy and is expressing their grief to a friend. We can imagine that what is important for them is not that the friend merely believes that they undergo a certain emotion but that the friend is in a much more intimate relation with their feelings. The friend may well know *that* they are grieving and *why* they are grieving, but what they want the friend to understand is *how* they really feel (Bonard & Deonna, 2023). And this may require that the friend undergoes, or at least has the capacity to undergo, affective states, which in turn requires being in certain phenomenal states. Radical affectivism is the theory that, for emotional communication to be optimally successful, the addressee must have the capacity to feel what the communicator expresses so as to understand *how* they feel (Bonard, 2021, Chapter 6).

If radical affectivism is correct, then phenomenal consciousness is required by at least one type of communication. Now, there may be other areas of discourse where phenomenal consciousness is required. Besides emotionally-loaded communication, intuitive candidates include communication using perceptual concepts (red, high-pitched, bitter) or other feeling concepts (feeling nauseous, tired, hungry). Furthermore, some may argue that grasping *any* concept requires phenomenal consciousness (e.g., Bourget, 2017; Pepperell, 2022). Such possibilities constitute massive obstacles for the MBM.

However, let me make four observations on behalf of the MBM. First, explaining phenomenal consciousness in functional terms may, in fact, be achievable, and this opens up the possibility of building AI with phenomenal consciousness (Butlin et al., 2023). Second, phenomenal consciousness, as opposed to access consciousness, is typically not required by theories of communication. Third, the view that *phenomenal* consciousness is required to grasp any kind of concept is, I gather, a minority view. Fourth, views such as radical affectivism are restricted to some forms of communication: even if they are correct, the MBM doesn't have to argue that AI will be able to master every form of communication.[11]

## 7. Conclusion

In the movie *Arrival*, extraterrestrial ships land on Earth with strange-looking, cephalopod-like aliens on board. Their intentions aren't clear. The US government asks a linguist, the main character of the movie, to decipher what these aliens may mean with the round stain-looking shapes they seem to communicate with. Much of the plot is about decoding these shapes. But there is a question that is not asked in the movie: Do these aliens genuinely communicate with these shapes, or is that just an illusion due to our tendency to anthropomorphize agent-looking creatures? That they genuinely communicate is presupposed in the plot, and as the movie progresses, it seems indeed reasonable to make this assumption. Readers who have seen the movie may ponder about what makes it seem indeed reasonable: Is it because they are living organisms? Because they seem to have their own goals and concerns? Because of the content of the messages they send?

I think comparing deep learning AIs with these aliens is illuminating insofar as, in both cases, humans are faced with unfamiliar black boxes (for a similar comparison, see Mitchell & Krakauer, 2023). Although current AIs have obvious communicative shortcomings (see sections 5 and 6 above), they may soon be like the aliens from *Arrival*: they will have all the appearances of genuine communicators. But they will furthermore use signs we already understand. For these reasons, and based on their intuitions, many people will probably consider that these future AIs can genuinely communicate with humans. But some skeptics will question it and ask for concrete evidence. My hope is that this paper indicates that the mental-behavioral methodology, together with progress in the cognitive science of communication, will help these skeptics answer their questions.[12]

---

11 See also Henry Shevlin's contribution 'Consciousness, machines, and moral status' in this volume for a discussion of the possibility of AI consciousness in light of the diversity of theories of consciousness.

12 Many thanks to Paula Droege for her comments on a previous version of this paper. Many thanks as well to Anna Strasser for her wonderful editorial work. This work was made possible thanks to the Swiss National Science Foundation grant P500PH_210771.